\documentclass[
 aip,
 jmp,
 amsmath,amssymb,
]{revtex4-2}

\usepackage{graphicx}
\usepackage{dcolumn}
\usepackage{bm}

\usepackage{float}
\usepackage{subcaption}
\usepackage{tabularx}
\usepackage{booktabs}
\usepackage{titlesec}
\usepackage{xspace}
\newcommand{\gao}{Ga\textsubscript{2}O\textsubscript{3}\xspace}
\newcommand{\sio}{SiO\textsubscript{2}\xspace}
\usepackage{comment}
\usepackage{makecell}
\usepackage{orcidlink}
\hypersetup{breaklinks=true}

\usepackage[a4paper,margin=2.5cm]{geometry}

\begin{document}

\preprint{}

\title[]{Study on the Effect of Annealing on \gao Thin Films Deposited on Silicon\\ by RF Sputtering}

\author{Ana Sofia Sousa\,\orcidlink{0009-0005-4018-8479}}
\email{ana.sofia.sousa@tecnico.ulisboa.pt}
\affiliation{INESC Microsystems and Nanotechnology, Rua Alves Redol 9, Lisbon 1000-029, Portugal}
\affiliation{Institute for Plasmas and Nuclear Fusion, Instituto Superior Técnico, University of Lisbon, Av. Rovisco Pais 1, Lisbon 1049-001, Portugal}

\author{Duarte M. Esteves\,\orcidlink{0000-0001-8566-9245}}
\affiliation{INESC Microsystems and Nanotechnology, Rua Alves Redol 9, Lisbon 1000-029, Portugal}
\affiliation{Institute for Plasmas and Nuclear Fusion, Instituto Superior Técnico, University of Lisbon, Av. Rovisco Pais 1, Lisbon 1049-001, Portugal}

\author{Tiago T. Robalo\,\orcidlink{0000-0001-6044-5333}}
\affiliation{Department of Physics, Faculty of Sciences, University of Lisbon, Campo Grande, Lisbon 1749-016, Portugal}
\affiliation{BioISI -- Biosystems and Integrative Sciences Institute, Faculty of Sciences, University of Lisbon, Campo Grande, Lisbon 1749-016, Portugal}

\author{M\'ario S. Rodrigues\,\orcidlink{0000-0002-0468-1910}}
\affiliation{Department of Physics, Faculty of Sciences, University of Lisbon, Campo Grande, Lisbon 1749-016, Portugal}
\affiliation{BioISI -- Biosystems and Integrative Sciences Institute, Faculty of Sciences, University of Lisbon, Campo Grande, Lisbon 1749-016, Portugal}

\author{Katharina Lorenz\,\orcidlink{0000-0001-5546-6922}}
\email{lorenz@ctn.tecnico.ulisboa@pt}
\affiliation{INESC Microsystems and Nanotechnology, Rua Alves Redol 9, Lisbon 1000-029, Portugal}
\affiliation{Institute for Plasmas and Nuclear Fusion, Instituto Superior Técnico, University of Lisbon, Av. Rovisco Pais 1, Lisbon 1049-001, Portugal}
\affiliation{Department of Nuclear Science and Engineering, Instituto Superior Técnico, University of Lisbon, Estrada Nacional 10, km 139.7, Bobadela 2695-066, Portugal}

\author{Marco Peres\,\orcidlink{0000-0001-6774-8492}}
\email{marcoperes@ctn.tecnico.ulisboa@pt}
\affiliation{INESC Microsystems and Nanotechnology, Rua Alves Redol 9, Lisbon 1000-029, Portugal}
\affiliation{Institute for Plasmas and Nuclear Fusion, Instituto Superior Técnico, University of Lisbon, Av. Rovisco Pais 1, Lisbon 1049-001, Portugal}
\affiliation{Department of Nuclear Science and Engineering, Instituto Superior Técnico, University of Lisbon, Estrada Nacional 10, km 139.7, Bobadela 2695-066, Portugal}

\begin{abstract}
Gallium oxide is an ultra-wide bandgap semiconductor with excellent opto-electronic properties, making it a highly promising material for a wide range of applications and devices. In this article, we report how the optical, morphological, structural, and compositional properties of $\beta$-\gao thin films deposited by RF sputtering on silicon substrates are affected by thermal treatments. Ellipsometric spectra recorded at multiple angles of incidence from several samples subjected to thermal annealing in the range of \mbox{550--1000 °C} were analyzed to extract the optical functions using appropriate multilayer models. This analysis is complemented by compositional, structural, and morphological characterization techniques. A significant increase of the refractive index was found after annealing at \mbox{1000 °C}, accompanied by a stark improvement in the samples' crystalline structure, as confirmed by complementary structural and compositional characterization techniques.
\end{abstract}


\maketitle

\section{Introduction}

Gallium oxide (\gao), an emerging wide-bandgap semiconductor, has become increasingly popular in recent years due to its outstanding opto-electronic properties such as high breakdown electric field of 8 MV/cm, ultra-wide bandgap of $\sim$ 4.85 eV at room temperature \cite{pearton_review_2018}, tunable electrical conductivity \cite{orita_preparation_2002}, high thermal and chemical stability \cite{kaur_strategic_2021} and excellent radiation hardness \cite{higashiwaki_gallium_2012}.

Some of the key applications of \gao are Schottky barrier diodes, field effect transistors and deep-UV (DUV) photodetectors \cite{higashiwaki_gallium_2012, liu_review_2019, xu_gallium_2019}. It is especially promising for a new generation of DUV photodetectors, due to its solar blind response as well as a tuneable bandgap, through alloying \cite{kaur_strategic_2021, xu_gallium_2019, liao_wide_2021, sousa_effect_2026}.

Regarding optical characteristics, and due to its wide bandgap which provides high transparency in the visible and UV regions, this semiconductor is promising for integrated photonic applications in the UV to near-infrared (NIR) spectral range  \cite{zhou_demonstration_2019, han_silicon-based_2025}. In addition to its potential for the development of low-loss waveguides, this material also stands out for its promise as a fast semiconductor scintillator, being considered the fastest among the semiconductor scintillators available for pulse height spectra \cite{li_enhanced_2022}. In this context, this semiconductor, which combines unique electrical and optical properties, is particularly interesting for the development of optoelectronic applications, and even for the development of Photonic Integrated Circuits (PICs) \cite{han_silicon-based_2025, lin_recent_2025}.

Furthermore, high-quality thin films of \gao can be obtained through epitaxial deposition, and bulk crystals can also be grown through cost-effective melt-growth processes \cite{jessen_supercharged_2021}. This work focuses on the potential of radio-frequency (RF) sputtered thin films, which is a fast technique requiring very simple equipment, resulting in a quality coating over large areas on a wide variety of substrates \cite{xu_gallium_2019}. By using thin films, we can leverage the micro- and nanofabrication techniques already developed and well-established for the silicon industry \cite{jessen_supercharged_2021}, as well as innovate through the use of flexible substrates, for wearable devices \cite{kaur_strategic_2021}.

However, despite its advantages, this physical deposition technique typically results in amorphous films of \gao when performed at room temperature, notwithstanding the crystalline structure of the substrate or gas conditions used \cite{ramana_properties_2019, patil_optical_2022}. Therefore, post-thermal treatments or performing the deposition at a higher temperature are essential for promoting the crystallization of these thin films \cite{ramana_properties_2019, mishra_effect_2022}. Despite their fundamental importance for the design of optical applications, the correlation between optical properties such as the refractive index and the structural, morphological, and compositional changes induced by thermal annealing remains insufficiently explored. In this work, we systematically investigate the effects of thermal annealing on the refractive index of \gao thin films deposited by RF sputtering on silicon substrates, together with the associated changes in structure and morphology.

\section{Materials and Methods}

The \gao thins films were deposited by RF sputtering at 60 W and 6 mTorr, in a home-built chamber at room temperature. The substrate holder was placed at approximately 7.3 cm above the 2$''$ \gao target (99.99\% purity), with an Ar flow of 15 sccm, for 1 h. A $\sim$ 3.6 cm $\times$ 3.6 cm piece of silicon wafer was used, which had been previously cleaned with acetone, isopropyl alcohol and deionized water.

The thickness of the thin film was determined through profilometry, performed on a Dektak XT, with a 2.5 \textmu m stylus and scan speed of 6.67 \textmu m/s. These measurements were done on a 3 by 3 grid.

The sample was then cut into 9 pieces, using a diamond-tipped pen. Following this, a study was performed by annealing them at different temperatures, from 550 to \mbox{1000 °C}, in \mbox{150 °C} steps, for 1 h. This was done in air, using a tube furnace.

Ellipsometry measurements were performed on an SE-2000 Spectroscopic Ellipsometer \cite{makai_breakthrough_2021}, with a spectral range spanning from 190 to 2100 nm, at incidence angles between 50 and \mbox{75°}, with a step of \mbox{5°}. The resulting data was processed using Semilab's SEA ellipsometric analysis software to perform the simulations and fits.

Atomic force microscopy (AFM) was done on a PicoLE Molecular Imaging AFM. These measurements were done in tapping mode, scanning a 1 \textmu m $\times$ 1 \textmu m area at a time, using a cantilever with a force constant of 5.4 N/m and a tip with a nominal radius smaller than 8 nm. The data was then processed using Gwyddion \cite{necas_gwyddion_2012}.

Rutherford backscattering spectrometry (RBS) was carried out, using the 2.5 MV Van de Graaff accelerator at Instituto Superior Técnico \cite{alves_insider_2021} to generate a 1.8 MeV He\textsuperscript{+} beam to probe the samples. Two Si pin diode detectors were used, with a resolution of 15 keV, placed at 165° and 140° with respect to the beam incident direction, and the samples were tilted by 60° towards the detectors in order to enhance the depth resolution. The data from both detectors was fitted simultaneously using NDF \cite{barradas_advanced_2008}, yielding information about the composition depth profile of the samples, while taking their surface roughness into account.

X-ray diffraction was done in order to probe the crystalline structure of the thin films. These scans were done on a Bruker D8 Discover Diffractometer, which has a Cu anode and W cathode, operated at 40 kV/40 mA, in low resolution mode. The primary beam was parallelized using a parabolic Göbel mirror and collimated using a 0.2 mm slit; the diffracted beam was collimated using long Soller slits and measured using a scintillation detector. Each sample was aligned using the $400$ reflection from the silicon substrate \cite{harrington_back--basics_2021}, which is the most intense peak, and the width of its $K_{\alpha1}$ peak was approximated to be the instrumental width for any calculations. The grazing incidence scans were taken at a 3° incidence angle, with the primary beam collimated by a 0.6 mm slit.

\section{Results \& Discussion}

The thickness of the as-deposited film considered for this study was determined to be 73 $\pm$ 4 nm by profilometry.

\begin{figure}[h]
\centering
\subfloat[\centering]{\includegraphics[width=0.48\textwidth]{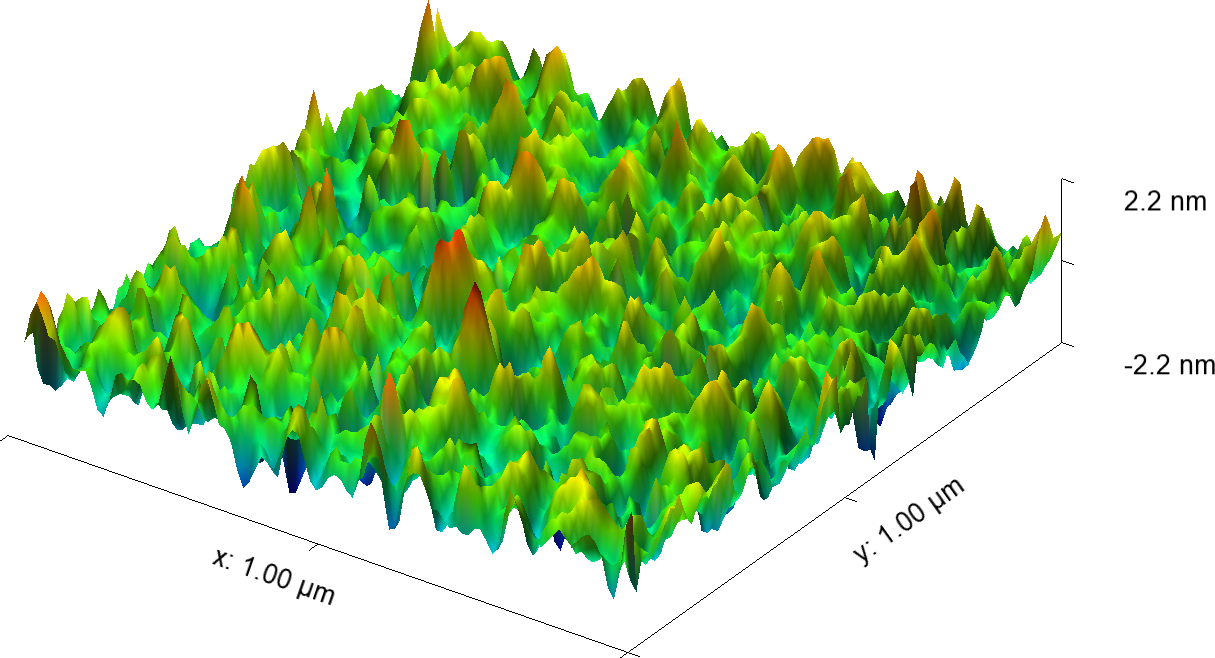}}
\hfill
\subfloat[\centering]{\includegraphics[width=0.48\textwidth]{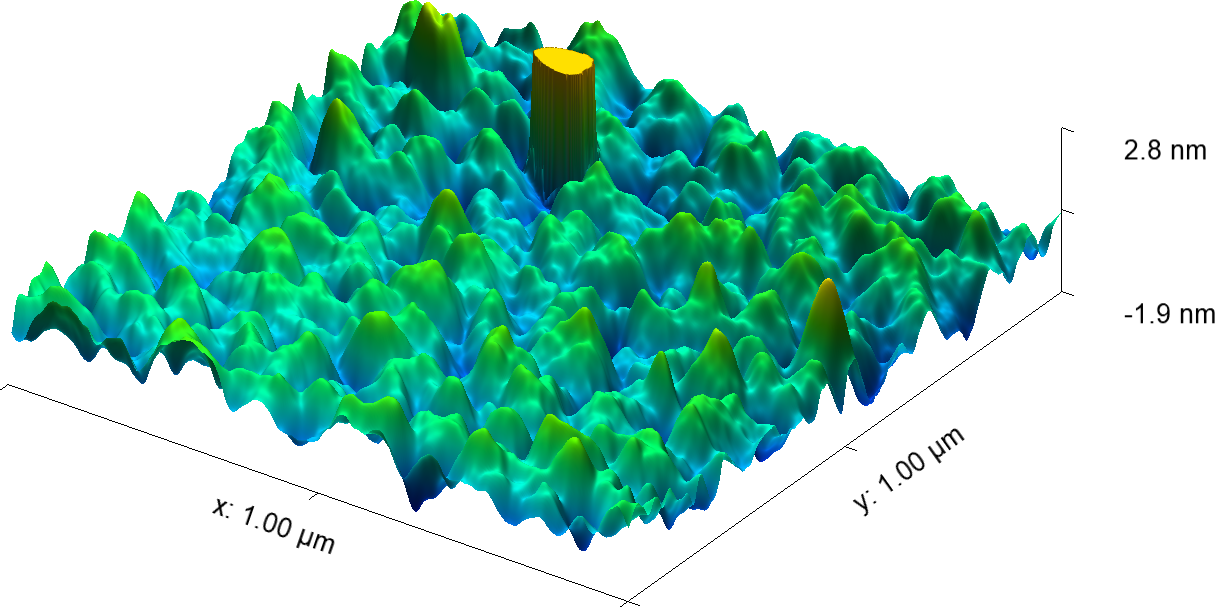}}
\hfill
\subfloat[\centering]{\includegraphics[width=0.48\textwidth]{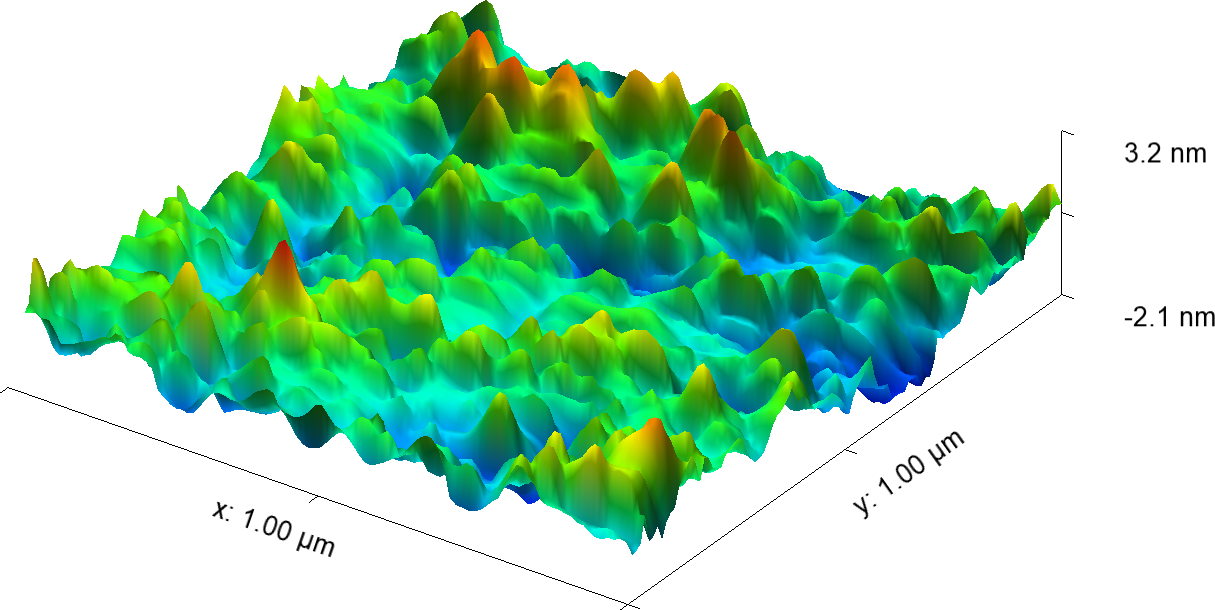}}
\hfill
\subfloat[\centering]{\includegraphics[width=0.48\textwidth]{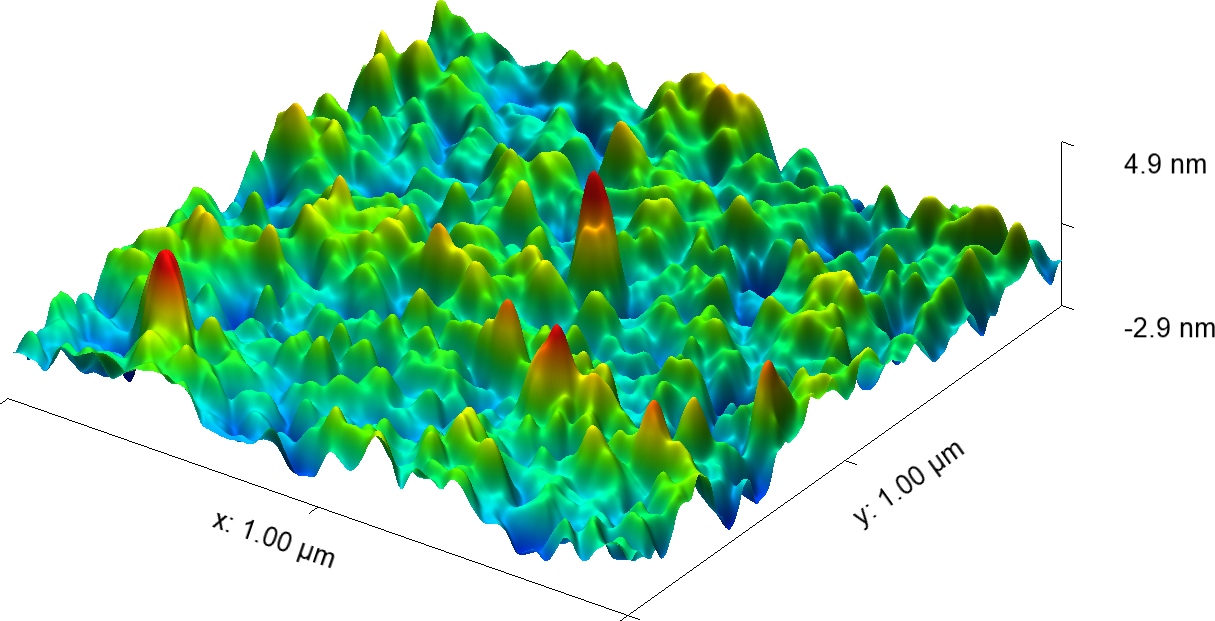}}
\hfill
\subfloat[\centering]{\includegraphics[width=0.48\textwidth]{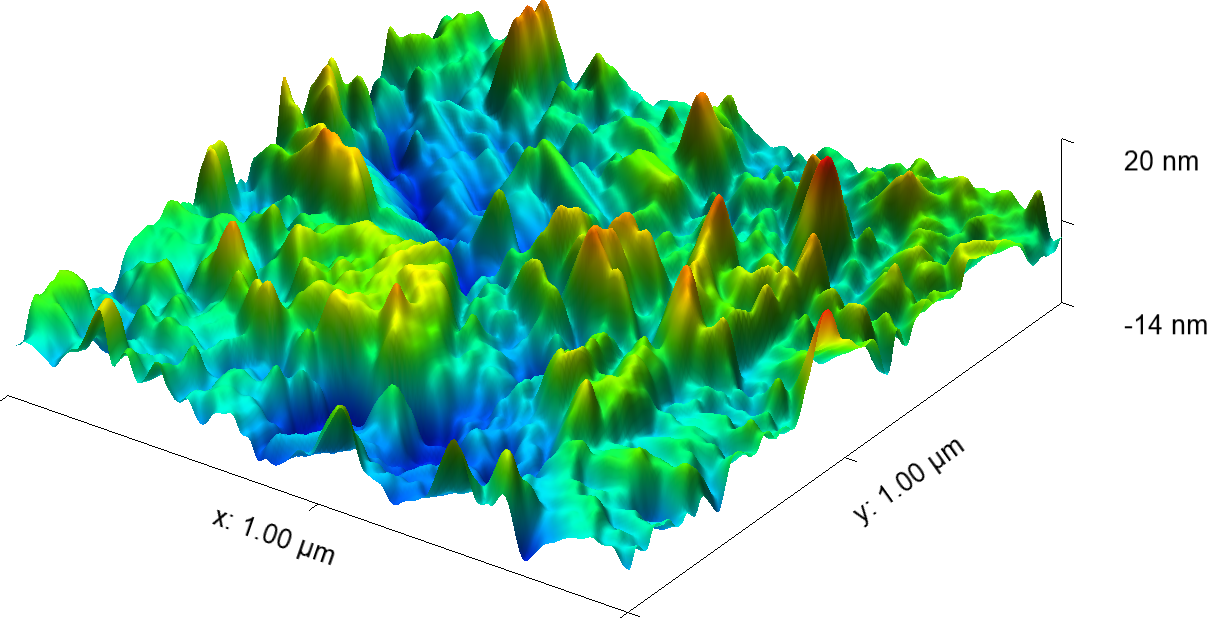}}
\caption{AFM images of the samples obtained (\textbf{a}) after deposition and after annealing at (\textbf{b}) 550, (\textbf{c}) 700, (\textbf{d}) 850, and (\textbf{e}) \mbox{1000 °C} .\label{afm}}
\end{figure}

Figure~\ref{afm} presents atomic force microscopy images of the as-deposited sample, as well as the samples annealed at various temperatures in air atmosphere. From these scans, the root mean square (RMS) roughness of the thin films was calculated, as shown in Table~\ref{afm-table}, demonstrating a marked increase after annealing at 1000 °C. Regarding the images themselves, they show that the films appear to be uniform in terms of features, although some inhomogeneities in the surface morphology begin to appear at higher temperatures.

\begin{table}[h]
\caption{Root-mean-square (RMS) roughness values for the AFM scans.\label{afm-table}}
\centering
\begin{tabular}{ccc}
\toprule
\textbf{Sample}	& \textbf{AFM RMS roughness}\\
\midrule
as-grown & 0.5 nm\\
550 °C & 0.6 nm \\
700 °C & 0.7 nm\\
850 °C & 0.9 nm\\
1000 °C & 5.4 nm\\
\bottomrule
\end{tabular}
\end{table}

\begin{figure}[h]
\centering
\includegraphics[width=0.8\textwidth]{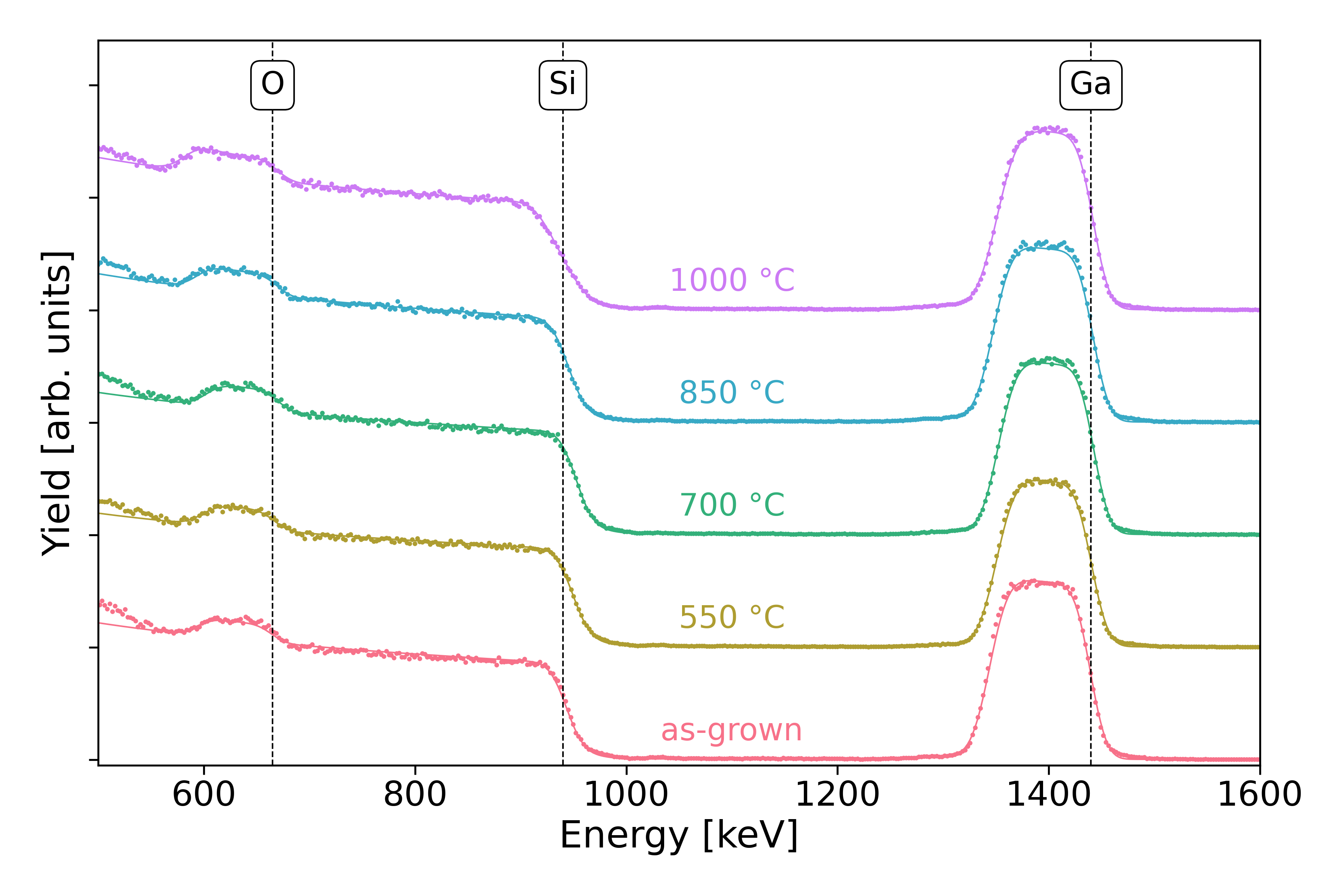}
\caption{RBS spectra and respective fits, for the samples as-grown and annealed at \mbox{550--1000 °C}. The energies corresponding to the Ga and O at the surface, as well as the Si barrier from the substrate/\sio, are marked. The data were vertically shifted for visual clarity and are from the detector placed at 165°.\label{rbs_spectra}}
\end{figure}

To determine their in-depth elemental composition and to evaluate the changes promoted by thermal treatment, the thin films were characterized using Rutherford backscattering spectrometry. As shown in Figure~\ref{rbs_spectra}, three distinct barriers are clearly observed in the RBS spectra, associated with the Ga and O from the thin film, in addition to the Si from the substrate. It is also worth noting that, with increasing temperature, no significant changes are observed that would suggest variations in composition or thickness. 

\begin{figure}[H]
    \centering
    \includegraphics[width=0.8\linewidth]{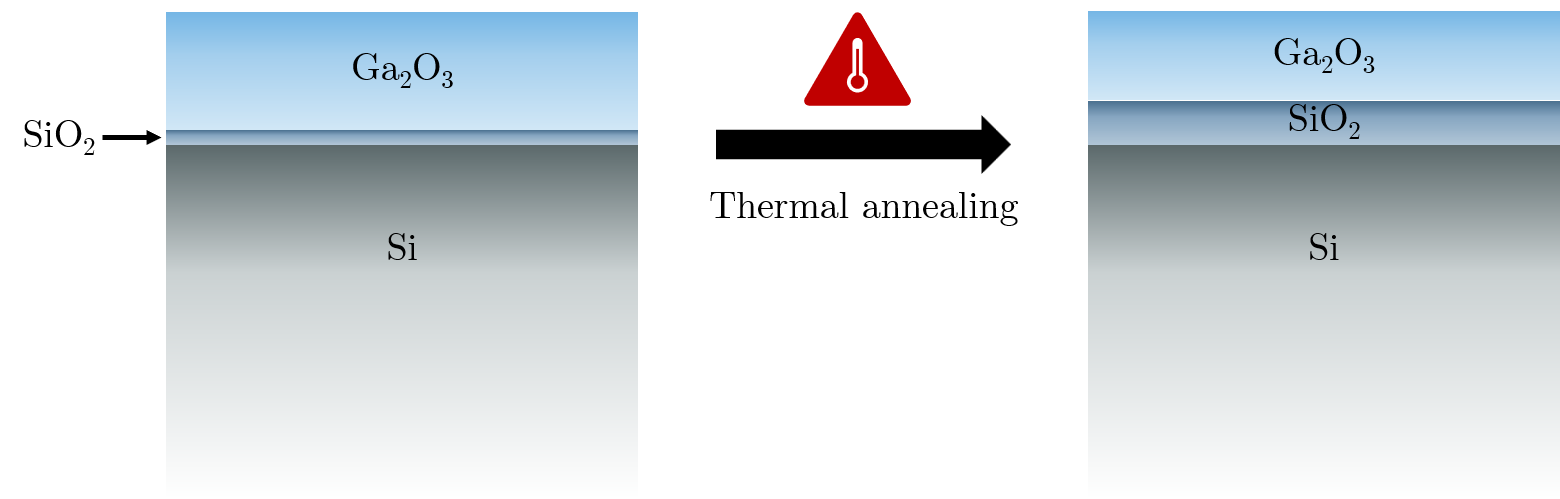}
    \caption{Schematic of the layer model used for the depth analysis.}
    \label{fig:layers}
\end{figure}

However, a more detailed analysis based on fitting these spectra indicated the presence of an \sio layer, as depicted in Figure~\ref{fig:layers}, whose thickness increases with annealing, especially at the highest temperature. The presence of this interface layer was expected, due to natural oxidation processes taking place before the sputtering \cite{morita_growth_1990, yen_role_2021} thickening due to the susceptibility of the Silicon substrate to oxidation \cite{altuntas_effect_2014}. It is noteworthy that this result is contrary to other studies, particularly those involving rapid thermal annealing (RTA), in which no alteration of the native \sio layer was observed as a consequence of the thermal annealing in air \cite{yen_role_2021}. This is a key characteristic of RTA, as the smaller annealing times prevent diffusion between layers \cite{muhammad_post-rapid_2025}.

Table~\ref{rbs_fit_results} presents the fitting results, showing that the thickness of the \gao layer tends to decrease slightly as the annealing temperature increases. On the other hand, an \sio layer is clearly present, and increasing in thickness with the annealing temperature. There is also a shift in the Ga/O stoichiometry, from 44/56\% to closer to 40/60\% after annealing. This effect was also observed by Gu \textit{et al.} \cite{gu_temperature-dependent_2022}, who found that oxygen could be diffused into extremely oxygen-deficient thin films deposited by atomic layer deposition through thermal annealing in ambient O\textsubscript{2}. This improvement in stoichiometry, which in this case suggests a reduction in oxygen vacancies, has been shown in some studies to significantly enhance the performance of \gao thin-film-based photodetectors \cite{feng_influence_2018, wang_role_2024}.  For the highest annealing temperature, surface roughness had to be considered to achieve a good fit, in good agreement with the AFM results in Table~\ref{afm-table}.

\begin{table}[H] 
\footnotesize 
\caption{RBS fit results, including layer thicknesses and stoichiometry.\label{rbs_fit_results}}
\centering
\begin{tabular}{c|ccccc}
\toprule
\textbf{Sample} 
& \makecell{\textbf{Ga\textsubscript{2}O\textsubscript{3} Thickness} \\ \textbf{\boldmath[$10^{15}$ at/cm$^2$]}} 
& \makecell{\textbf{Roughness} \\ \textbf{\boldmath[$10^{15}$ at/cm$^2$]}} 
& \textbf{Ga [\%]} 
& \textbf{O [\%]} 
& \makecell{\textbf{SiO\textsubscript{2} Thickness} \\ \textbf{\boldmath[$10^{15}$ at/cm$^2$]}} \\
\midrule
as-grown & 594 & 0 & 44 & 56 & 6 \\
550 °C & 585 & 0 & 39 & 61 & 15 \\
700 °C & 576 & 0 & 39 & 61 & 24 \\
850 °C & 571 & 0 & 40 & 60 & 62 \\
1000 °C & 537 & 40 & 40 & 60 & 202 \\
\bottomrule
\end{tabular}
\end{table}

\begin{figure}[H]
\centering
\includegraphics[width=0.64\textwidth]{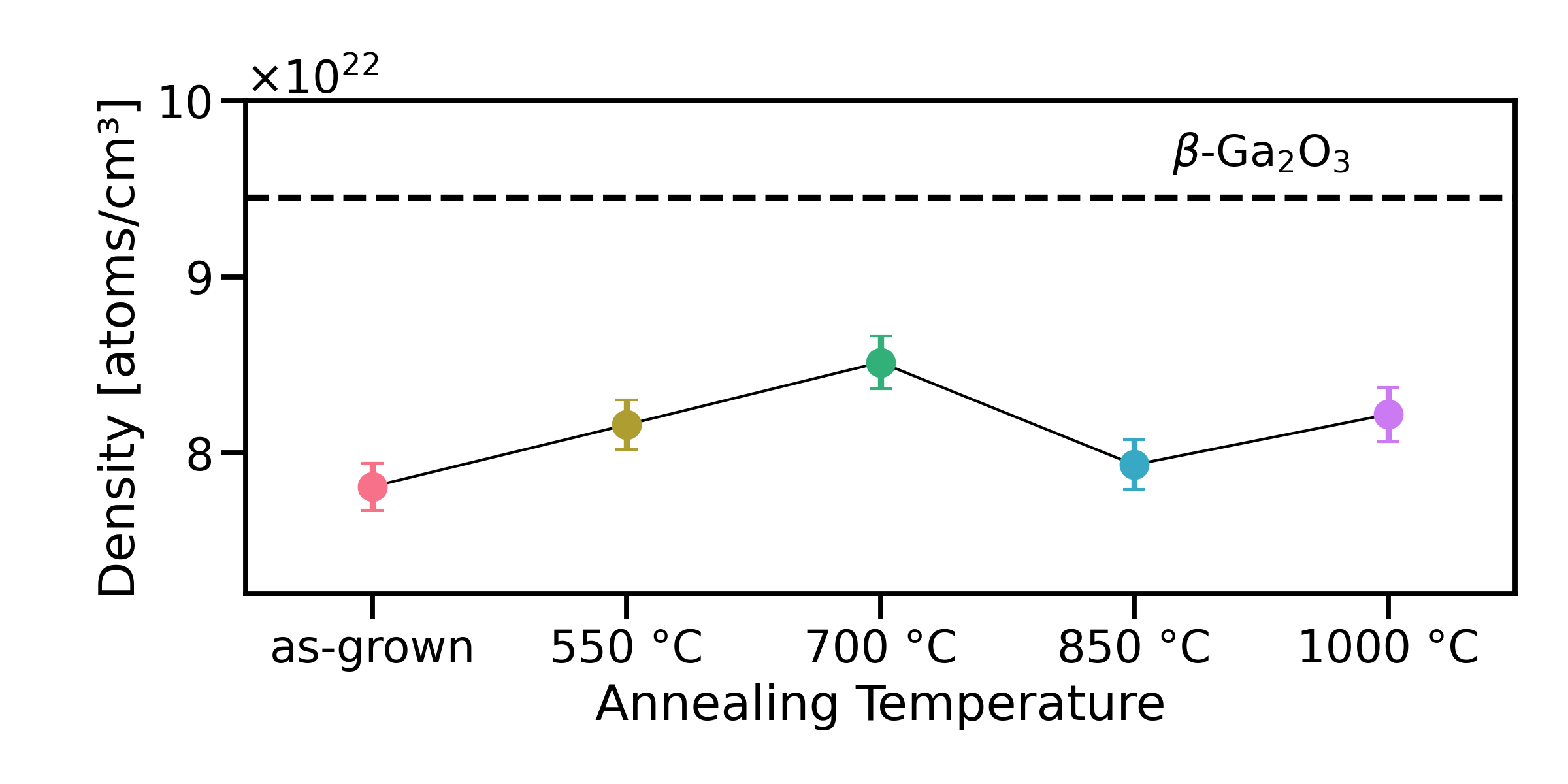}
\caption{Density of the thin films, estimated from the RBS and ellipsometry thicknesses in Tables~\ref{rbs_fit_results} and \ref{ellipso}, respectively.\label{density}}
\end{figure}

These results, in conjunction with the films' thicknesses as determined through ellipsometry (Table~\ref{ellipso}) allow us to estimate their densities. As shown in Figure~\ref{density}, the density tends to increase as the annealing temperature increases, approaching the theoretical atomic density of \mbox{9.45 $\cdot$ 10\textsuperscript{22} atoms/cm\textsuperscript{3}}  \cite{stepanov_gallium_2016} for $\beta$-\gao.

\begin{figure}[h]
\centering
\includegraphics[width=0.8\textwidth]{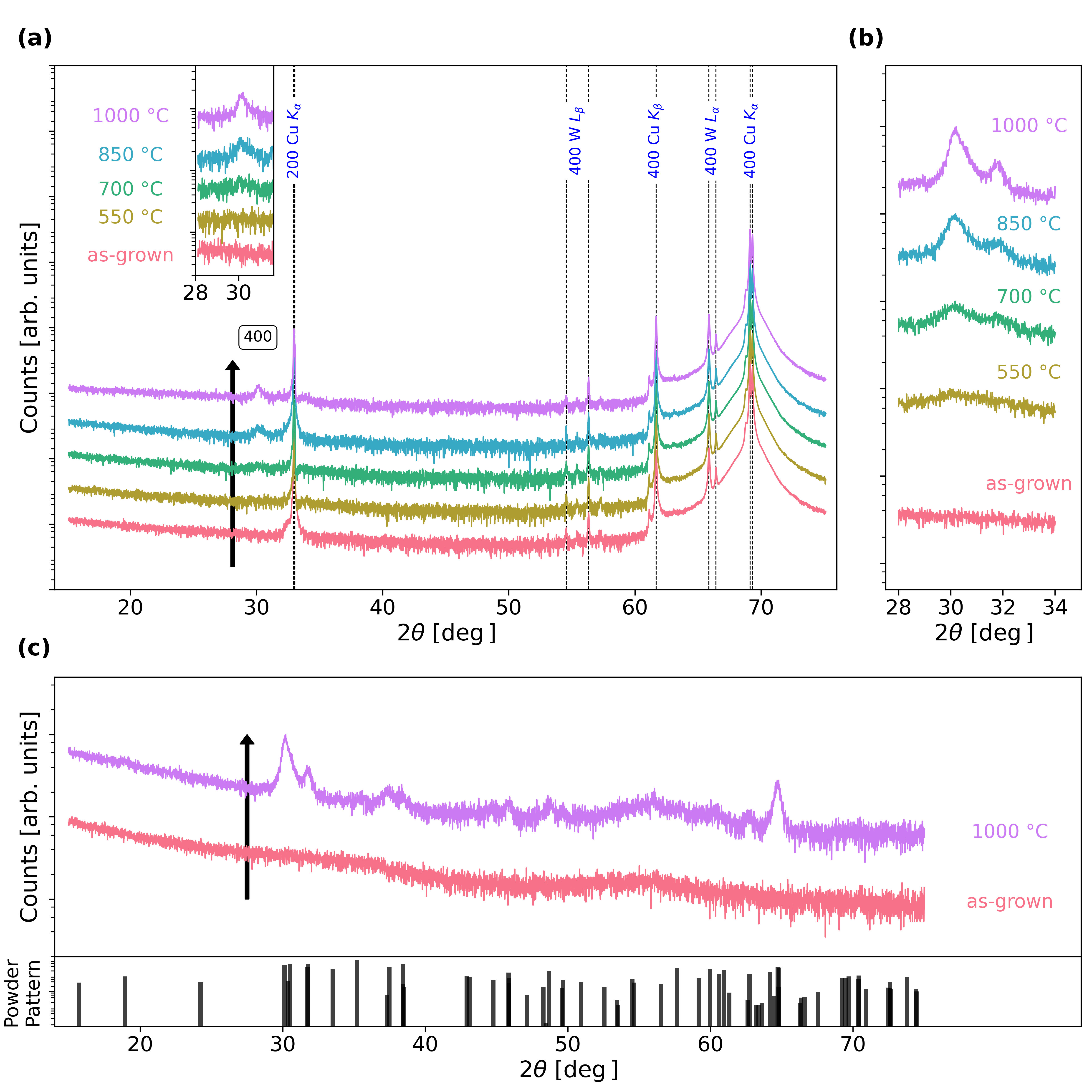}

\caption{X-Ray diffractograms of the thin films as a function of annealing temperature, (\textbf{a}) in a 2$\theta$-$\omega$ scan with the most relevant substrate peaks marked in blue as well as the $400$ \gao reflection, (\textbf{b}) in a grazing incidence scan around the $400$ reflection, and (\textbf{c}) showing the full grazing incidence scan and the powder pattern \cite{ahman_reinvestigation_1996}. The data were vertically shifted for visual clarity.\label{xrd_diffs}}
\end{figure}

To complement the compositional analysis, a structural characterization of these films was performed using X-ray diffraction (XRD). Figure~\ref{xrd_diffs}(\textbf{a}) features the XRD $2\theta$--$\omega$ scans, showing no peaks in the as-deposited films, suggesting that they are amorphous as expected for room-temperature deposition, and the emergence of a peak at around 30° as the annealing temperature increases, tentatively assigned to the $400$ reflection of $\beta$-\gao. To better ascertain the crystalline quality of the thin films, grazing incidence scans were then taken --- Figures~\ref{xrd_diffs}(\textbf{b}) and \ref{xrd_diffs}(\textbf{c}). Figure~\ref{xrd_diffs}(\textbf{b}) shows an improvement in crystalline quality with annealing temperature and figure~\ref{xrd_diffs}(\textbf{c}) clarifies that our film is textured, as a multitude of peaks can be observed in this diffractogram, with intensity ratios differing from the powder pattern. This is congruent with previous studies --- Makeswaran \textit{et al.} \cite{makeswaran_crystal_2020} also observed a variety of peaks, that became more intense and defined with increasing annealing temperature.

\begin{figure}[H]
\centering
\includegraphics[width=0.8\textwidth]{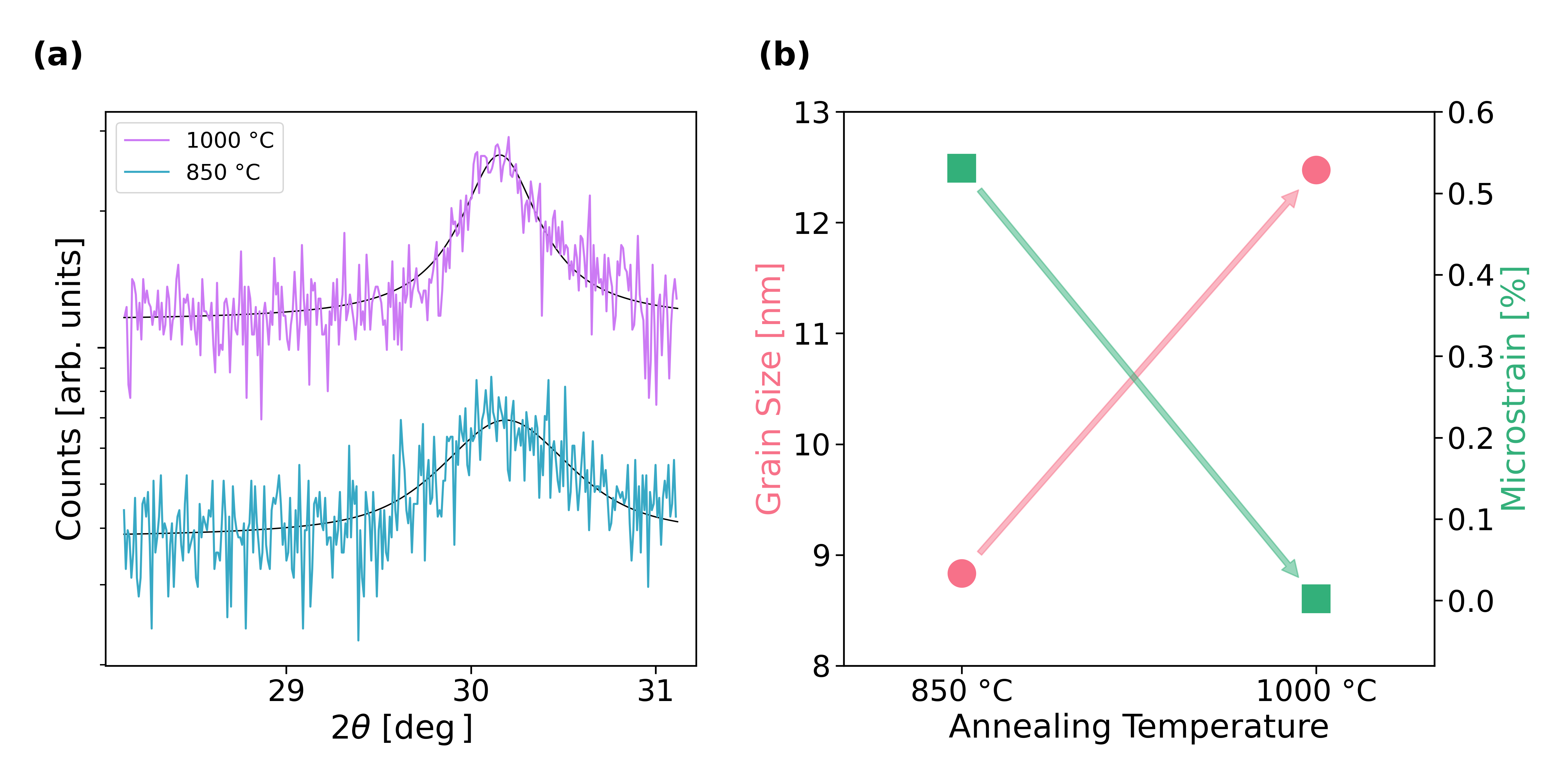}
\caption{Analysis of the $400$ peaks of the thin-films annealed at 850 and 1000 °C: (\textbf{a}) pseudo-Voigt fitting, and (\textbf{b}) corresponding grain size and microstrain values.\label{xrd_analysis}}
\end{figure}

We can leverage this data to obtain further information on the crystallinity of the samples. By fitting the $\beta$-\gao $400$ peak to a pseudo-Voigt function, as shown in Figure~\ref{xrd_analysis}(\textbf{a}), we can estimate the crystal domain size and microstrain through the Scherrer \cite{scherrer_bestimmung_1918} and Wilson \cite{stokes_diffraction_1944} equations, respectively. We opted to do this only for the samples annealed at 850 and 1000 °C, as these showed clearly defined peaks. Figure~\ref{xrd_analysis}(\textbf{b}) then shows the evolution of these two parameters, demonstrating an increase in the grain size, as expected, as well as a dramatic decrease in the microstrain \cite{sinha_effect_2007}. This conclusively demonstrates that the crystallinity of the thin film improves with annealing temperature, in agreement with other studies \cite{gu_influence_2024}. On the other hand, these results also suggest that the increase in roughness measured by both AFM and RBS is likely associated with structural improvement and the formation of larger grains \cite{guo_growth_2014, dong_effects_2016, gu_influence_2024}.

To study and characterize the optical properties of the thin films, variable angle spectroscopic ellipsometry was used. Based on this method and through fitting the obtained results, it was possible to estimate the refractive index and the extinction coefficient of the deposited films, as well as their thicknesses. As with the RBS analysis, the model was based on three layers as shown on Figure~\ref{fig:layers}, using a Tauc-Lorentz dispersion function \cite{jellison_parameterization_1996, jellison_erratum_1996} to model the thin \gao film.

\begin{figure}[H]
\centering
\includegraphics[width=0.8\textwidth]{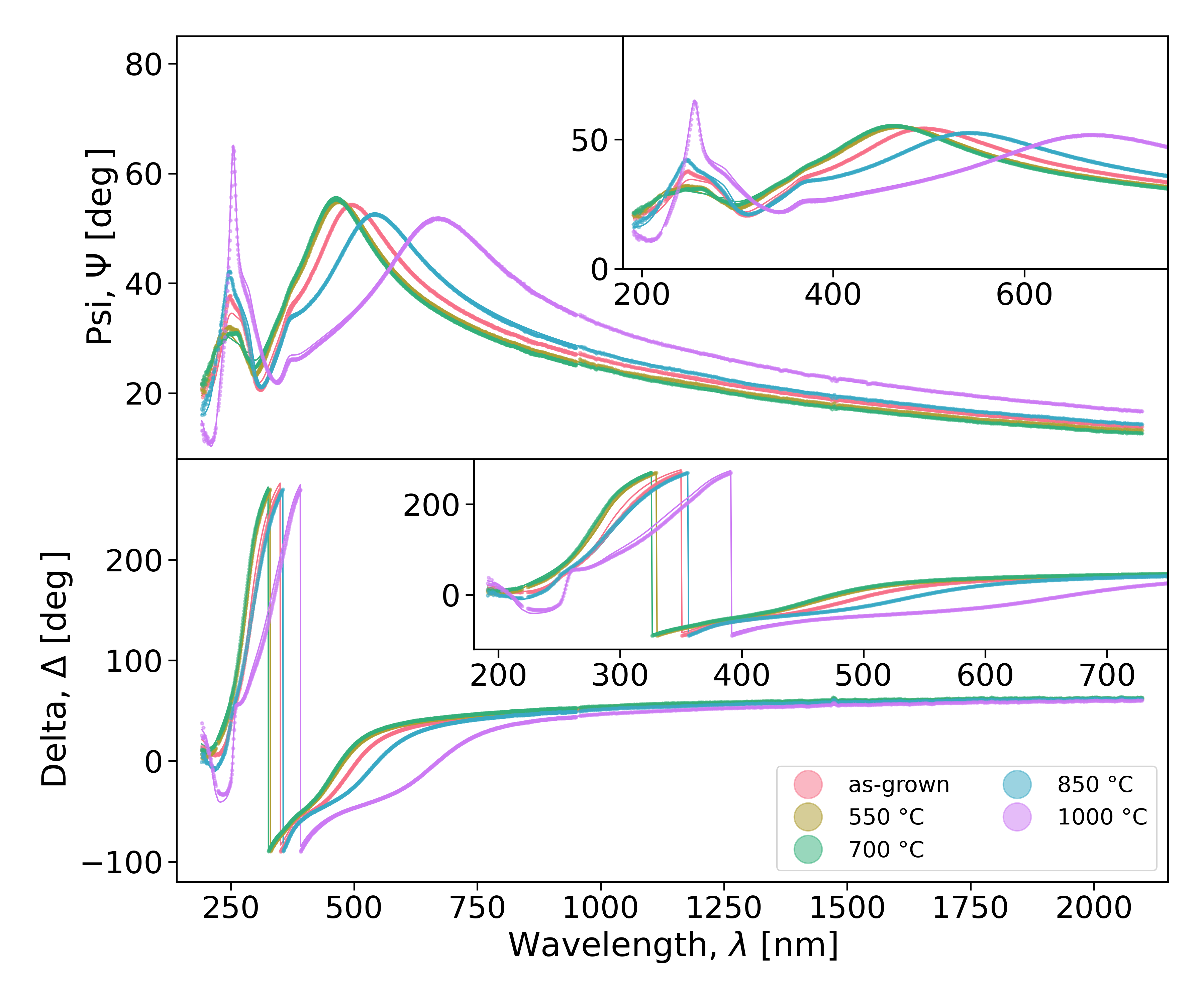}
\caption{Ellipsometric Psi ($\Psi$) and Delta ($\Delta$) spectra and fits of the as-grown and annealed samples, taken at 75°. For clarity, there are insets showing the data at wavelengths 190--750 nm. \label{psi-delta}}
\end{figure}

Figure~\ref{psi-delta} shows the spectra acquired at 75° along with the fittings, as a function of the annealing temperature. It shows that the model is quite adequate in fitting the data, as further demonstrated by the high coefficient of determination $R^2$ values in Table~\ref{ellipso}, as well as taking into account the expected thickness of 73 nm. This table shows the different Tauc-Lorentz (TL) parameters: $A$ is the oscillator amplitude, $E_\text{0}$ is the oscillator peak position, $C$ represents the peak width and $E_\text{g}$ is the bandgap. This table outlines how the layer of \sio thickens with annealing. Although this technique is more sensitive to minute variations in thickness than RBS, we found that it could not reliably fit the roughness of the thin film. This is unsurprising, as we are considering a large number of variable parameters which may have indistinguishable effects while fitting.

\begin{table*}[h] 
\scriptsize
\caption{Thickness of the \gao and \sio layers, bandgap, refractive index (at 632.8 nm) and Tauc-Lorenz (TL) parameters resulting from the ellipsometry fit.\label{ellipso}}
\begin{tabular}{c|ccccccccc}
\toprule
\textbf{Sample} & $\boldsymbol{n}$ & \makecell{\textbf{Ga\textsubscript{2}O\textsubscript{3}}\\ \textbf{thickness [nm]}} & \makecell{\textbf{SiO\textsubscript{2}}\\ \textbf{thickness [nm]}} & \textbf{TL} $\boldsymbol{A}$ \textbf{[eV]} & \textbf{TL} $\boldsymbol{E_{\text{0}}}$ \textbf{[eV]} & \textbf{TL} $\boldsymbol{C}$ \textbf{[eV]} & \textbf{TL} $\boldsymbol{E_{\text{g}}}$ \textbf{[eV]} & \textbf{TL} $\boldsymbol{\epsilon}_{\mathbf{\infty}}$ & $\boldsymbol{R^2}$ \\
\midrule
as-grown & 1.851 & 76.1 $\pm$ 0.3 & 2.6 $\pm$ 0.2 & 130 $\pm$ 6 & 5.28 $\pm$ 0.08 & 5.0 $\pm$ 0.1 & 4.17 $\pm$ 0.02 & 2.06 $\pm$ 0.03 & 0.998 \\
550 °C & 1.860 & 71.7 $\pm$ 0.3 & 2.0 $\pm$ 0.2 & 112 $\pm$ 6 & 6.59 $\pm$ 0.06 & 8.7 $\pm$ 0.4 & 3.97 $\pm$ 0.02 & 1.80 $\pm$ 0.04 & 0.999 \\
700 °C & 1.863 & 67.6 $\pm$ 0.2 & 4.8 $\pm$ 0.2 & 120 $\pm$ 5 & 5.87 $\pm$ 0.06 & 6.5 $\pm$ 0.2 & 4.110 $\pm$ 0.009 & 1.97 $\pm$ 0.03 & 0.999 \\
850 °C & 1.865 & 72.0 $\pm$ 0.3 & 12.8 $\pm$ 0.2 & 153 $\pm$ 8 & 5.7 $\pm$ 0.2 & 8.2 $\pm$ 0.2 & 4.13 $\pm$ 0.02 & 1.76 $\pm$ 0.03 & 0.998 \\
1000 °C & 1.915 & 65.3 $\pm$ 0.2 & 37.3 $\pm$ 0.1 & 138 $\pm$ 4 & 5.62 $\pm$ 0.04 & 4.68 $\pm$ 0.04 & 4.280 $\pm$ 0.008 & 2.14 $\pm$ 0.02 & 0.999 \\
\bottomrule
\end{tabular}
\end{table*}

The bandgap seems to slightly decrease after the initial annealing at \mbox{550 °C}, and then increases again after annealing at \mbox{700--1000 °C}. These changes are aligned with the results reported by Muhammad \textit{et al.} \cite{muhammad_tailoring_2025}, and correlate nicely with the RBS stoichiometry results --- with annealing, the amount of oxygen vacancies decreases and the bandgap increases \cite{lim_crystallization_2022}.

\begin{figure}[h]
\centering
\includegraphics[width=\textwidth]{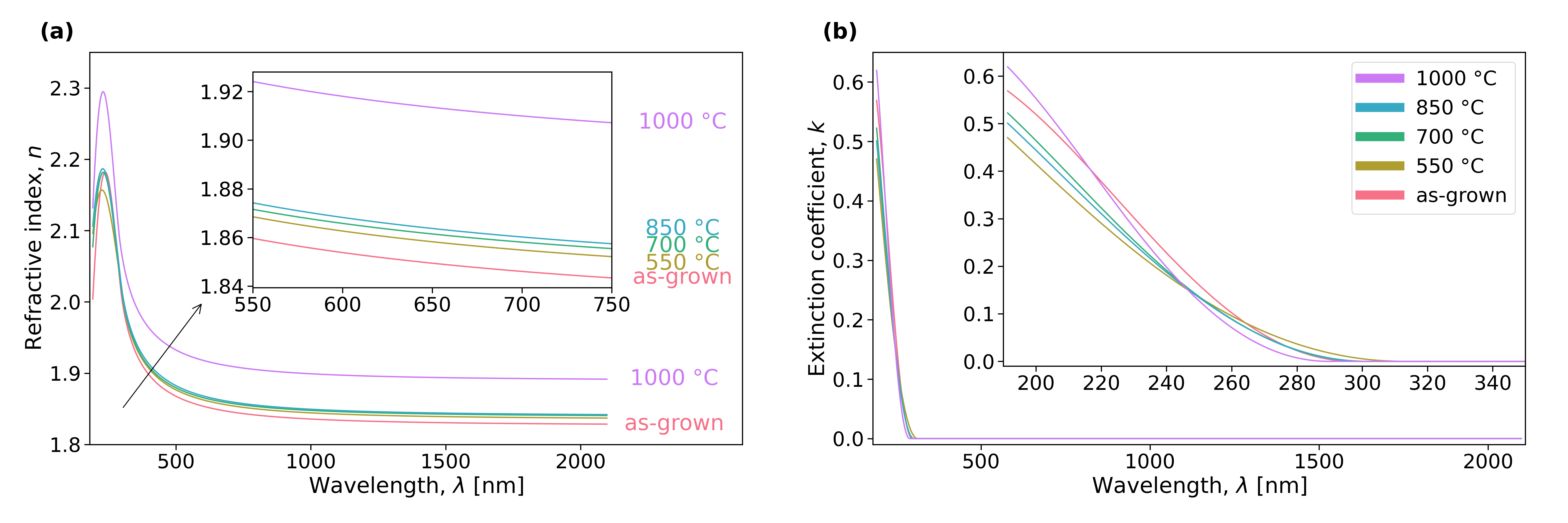}
\caption{Spectra of the (\textbf{a}) refractive index ($n$) and (\textbf{b}) extinction coefficient ($k$) taken from the TL model. For clarity, there are insets showing the data at wavelengths 550--750 and 190--350 nm, respectively.\label{refractive_index}}
\end{figure}

Figure~\ref{refractive_index}(\textbf{a}) shows the resulting refractive index spectra, outlining an increase of the refractive index with annealing temperature, especially after annealing at 1000 °C. This is congruent with results in literature, as Makeswaran \textit{et al.} observed a similar increase in samples annealed at \mbox{500--900 °C} \cite{makeswaran_crystal_2020}. It is also in agreement with the overall increase in density shown in Figure~\ref{density}, as an increase in the refractive index of a material generally indicates denser films \cite{altuntas_effect_2014, tan_dependence_2021}. The extinction coefficient $k$, in Figure~\ref{refractive_index}(\textbf{b}), is null up to roughly 300 nm, demonstrating excellent transparency in the UVA--visible--NIR spectral region. This onset of extinction is in agreement with the material's bandgap.

These interesting results warranted further investigation, which can be done by modeling these same data with the Wemple-DiDomenico (WDD) single-effective oscillator model \cite{keerthana_c_s_understanding_2024, wemple_behavior_1971}. This simple analysis, stemming from the refractive index information obtained through ellipsometry, yields three key parameters: $E_{\text{0}}$, or the single oscillator energy, which gives quantitative information on the overall band structure of the material; $E_{\text{d}}$, or the dispersion energy, which measures the average strength of the interband optical transitions; and $n_{\text{0}}$, or the static refractive index, which is related to the electric polarizability and optical bandgap of the material \cite{borah_investigation_2020, wemple_behavior_1971, gomaa_correlation_2021}.

\begin{figure}[H]
\centering
\includegraphics[width=0.8\textwidth]{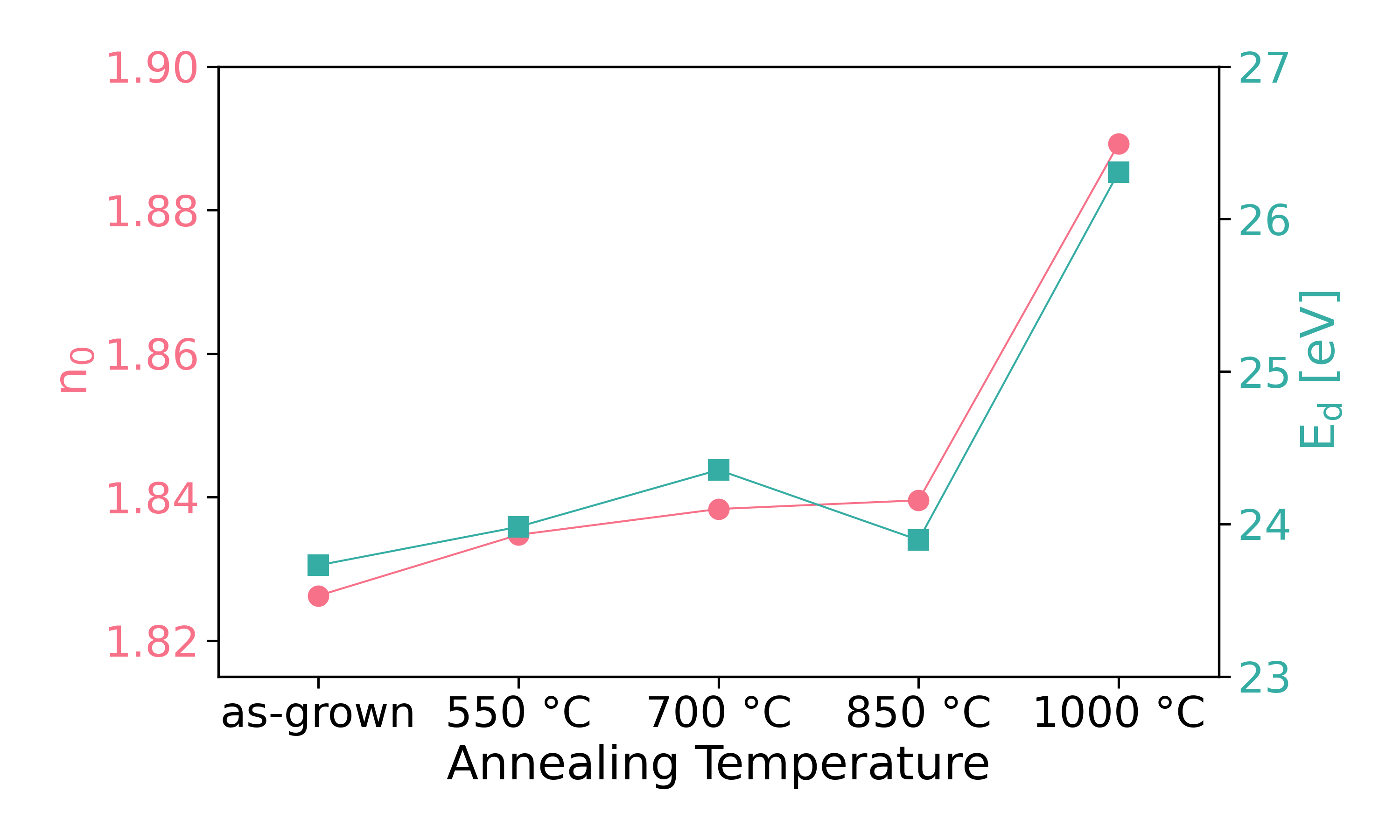}
\caption{Static refractive index $n_{\text{0}}$ and dispersion energy $E_{\text{d}}$ resulting from the Wemple-DiDomenico analysis of the thin films, from the ellipsometry data.\label{wdd}}
\end{figure}

Thin films are complex systems which are subject to point defects or imperfections in the crystal lattice, such as vacancies and interstitials, but also to the effect of grain boundaries. These form bands within the bandgap, which do not have a significant impact on the oscillator energy, but do affect the material's response to different frequencies of light \cite{borah_investigation_2020}. As shown in Figure~\ref{wdd}, thermal annealing leads to an increase of $n_{\text{0}}$ and $E_{\text{d}}$, which is more marked after the final \mbox{1000 °C} annealings. As reported by Borah \textit{et al.} \cite{borah_investigation_2020}, this is due to a reduction of structural disorder, which is in agreement with our XRD results, and can lead to an enhancement of the optical response of the thin films.

\section{Conclusions}

Stemming from the analysis done through four complementary techniques --- AFM, RBS, XRD and ellipsometry  ---, we clearly demonstrated a shift in the structural and optical properties of RF-sputtered thin films with thermal annealing. It was shown that there is an increase of the refractive index with the increase of annealing temperature, accompanied by the thickening of an \sio layer, observed both through RBS and ellipsometry. We saw a definite improvement on the crystalline quality of the samples through XRD, marked by increasing grain size and decreasing microstrain, supported by an increase in the density of the films. Direct measurements of the samples' roughness by AFM showed a very significant increase after annealing at \mbox{1000 °C}. This study is of special relevance for the understanding of the optical properties of this material and how they can potentially be tuned for applications such as waveguides, as well as for the integration of electrical and optical functionality in the same device.

\section*{Author Contributions}

Conceptualization, A.S.S., K.L. and M.P.; methodology, A.S.S., D.M.E., T.T.R., M.S.R., K.L. and M.P.; software, A.S.S.; validation, K.L. and M.P.; formal analysis, A.S.S.; investigation, A.S.S., D.M.E., T.T.R., M.S.R., K.L. and M.P.; resources, M.S.R., K.L. and M.P.; data curation, A.S.S. and T.T.R.; writing---original draft preparation, A.S.S.; writing---review and editing, A.S.S., D.M.E., T.T.R, M.S.R, K.L. and M.P.; visualization, A.S.S.; supervision, M.S.R., K.L. and M.P.; project administration, K.L. and M.P.; funding acquisition, K.L. and M.P.. All authors have read and agreed to the published version of the manuscript.

\section*{Acknowledgments}

The authors acknowledge funding of the Research Unit INESC MN from the Fundação para a Ciência e a Tecnologia (FCT) through Plurianual financing (UIDB/05367/2025, UID/PRR/5367/2025 and UID/PRR2/05367/2025) [doi: \linebreak https://doi.org/10.54499/UIS/PRR/05367/2025 and \linebreak https://doi.org/10.54499/UID/PRR2/05367/2025], as well as via the IonProGO project (2022.05329.PTDC, http://doi.org/10.54499/2022.05329.PTDC) and via the INESC MN Research Unit funding (UID/05367/2020) through Pluriannual BASE and PROGRAMATICO financing.  A. S. Sousa thanks FCT for her PhD grant (2025.04778.BD), as does D. M. Esteves (2022.09585.BD, https://doi.org/10.54499/2022.09585.BD).

\section*{Data availability}

The raw data supporting the conclusions of this article will be made available by the authors on request.

\section*{Conflicts of interest}

The authors declare no conflicts of interest. The funders had no role in the design of the study; in the collection, analyses, or interpretation of data; in the writing of the manuscript; or in the decision to publish the results.

\bibliography{Bib}

\end{document}